\documentclass{elsart}

\usepackage{graphicx}
\usepackage{amssymb}

\begin{document}
\begin{frontmatter}

\title{Incommensurate magnetic response in cuprate perovskites}

\author{A.~Sherman}
\ead{alexei@fi.tartu.ee}

\address{Institute of Physics, University of Tartu, Riia 142, 51014
Tartu, Estonia}

\begin{abstract}
The incommensurate magnetic response of the normal-state cuprate
perovskites is interpreted based on Mori's memory function approach and
the $t$-$J$ model of Cu-O planes. In agreement with experiment the
calculated dispersion of the susceptibility maxima has the shape of two
parabolas with upward and downward branches which converge at the
antiferromagnetic wave vector. The maxima are located at
$(\frac{1}{2},\frac{1}{2}\pm\delta)$,
$(\frac{1}{2}\pm\delta,\frac{1}{2})$ and at
$(\frac{1}{2}\pm\delta,\frac{1}{2}\pm\delta)$,
$(\frac{1}{2}\pm\delta,\frac{1}{2}\mp\delta)$ in the lower and upper
parabolas, respectively. The upper parabola reflects the dispersion of
magnetic excitations of the localized Cu spins, while the lower
parabola arises due to a dip in the spin-excitation damping at the
antiferromagnetic wave vector. For moderate doping this dip stems from
the weakness of the interaction between the spin excitations and holes
near the hot spots.
\end{abstract}

\end{frontmatter}

One of the most interesting features of the inelastic neutron
scattering in lanthanum cuprates is that for hole concentrations $x
\gtrsim 0.04$, low temperatures and small energy transfers the
scattering is peaked at incommensurate momenta
$(\frac{1}{2},\frac{1}{2}\pm\delta)$,
$(\frac{1}{2}\pm\delta,\frac{1}{2})$ in the reciprocal lattice units
$2\pi/a$ with the lattice period $a$ \cite{Yoshizawa}. For $x\lesssim
0.12$ the incommensurability parameter $\delta$ is approximately equal
to $x$ and saturates for larger $x$ \cite{Yamada}. The incommensurate
response was observed both below and above $T_c$ \cite{Mason93}.
Recently the analogous low-frequency incommensurability was observed in
YBa$_2$Cu$_3$O$_{7-y}$ \cite{Dai}. This gives grounds to suppose that
the incommensurability is a common feature of cuprate perovskites which
does not depend on subtle details of the energy structure. However, for
larger frequencies the susceptibility differs essentially in these two
types of cuprates. In the underdoped YBa$_2$Cu$_3$O$_{7-y}$ both below
and above $T_c$ a maximum was observed at frequencies
$\omega_r=25-40$~meV \cite{Bourges}. For this frequency the magnetic
response is sharply peaked at the antiferromagnetic wave vector ${\bf
Q}=(\frac{1}{2},\frac{1}{2})$. Contrastingly, no maximum at $\omega_r$
was observed in lanthanum cuprates \cite{Aeppli}. For even larger
frequencies the magnetic response becomes again incommensurate in both
types of cuprates with peaks located at
$(\frac{1}{2}\pm\delta,\frac{1}{2}\pm\delta)$,
$(\frac{1}{2}\pm\delta,\frac{1}{2}\mp\delta)$
\cite{Bourges,Hayden04,Tranquada}. The dispersion of the maxima in the
susceptibility resembles two parabolas with upward and downward
branches which converge at ${\bf k=Q}$ and $\omega\approx\omega_r$
\cite{Tranquada}.

The nature of the magnetic incommensurability is the subject of active
discussion nowadays. The most frequently used approaches for its
explanation are based on the picture of itinerant electrons with the
susceptibility calculated in the random phase approximation \cite{Liu}
and on the stripe domain picture \cite{Tranquada,Hizhnyakov}. In the
former approach the low-frequency incommensurability is connected with
the Fermi surface nesting which exists in the normal state or arises
after the superconducting transition. In view of the similarity of the
incommensurability in yttrium and lanthanum cuprates this approach
implies the near resemblance of their Fermi surfaces and their change
with doping which is highly improbable. Besides, the applicability of
the picture of itinerant electrons for underdoped cuprates casts
doubts. As for the second notion, it should be noted that in the
elastic neutron scattering the charge-density wave connected with
stripes is observed only in crystals with the low-temperature
tetragonal or the low-temperature less-orthorhombic phases
(La$_{2-x}$Ba$_x$CuO$_4$ and La$_{2-y-x}$Nd$_y$Sr$_x$CuO$_4$) and does
not observed in La$_{2-x}$Sr$_x$CuO$_4$ in the low-temperature
orthorhombic phase \cite{Kimura}. At the same time the magnetic
incommensurability is similar in these two types of crystals. It can be
supposed that the magnetic incommensurability is the cause rather than
the effect of stripes which are formed with an assistance of phonons.

In the present work the general formula for the magnetic susceptibility
derived in the memory function approach \cite{Mori} is used. For the
description of spin excitations in the doped antiferromagnet the
$t$-$J$ model of a Cu-O plane is employed. In this approach the
mentioned peculiarities of the magnetic properties of cuprates are
reproduced including the proper frequency and momentum location of the
susceptibility maxima. The incommensurability for $\omega>\omega_r$ is
connected with the dispersion of spin excitations. The
incommensurability for lower frequencies is related to the dip in the
spin-excitation damping at ${\bf Q}$. For moderate doping this dip
stems from the weakness of the interaction between the spin excitations
and holes near the hot spots. Such form of the interaction constant
follows from the fact that a decaying site spin excitation creates a
fermion pair with components residing on the same and neighbor sites.

In the memory function approach the imaginary part of the magnetic
susceptibility, which is directly connected with the cross-section of
the magnetic scattering, is described by the formula \cite{Sherman02}
\begin{equation}\label{chi}
\chi''({\bf k}\omega)=-\frac{4\mu^2_B\omega\Im R({\bf k}\omega)
}{[\omega^2-\omega f_{\bf k}\Re R({\bf k}\omega)-\omega^2_{\bf
k}]^2+[\omega f_{\bf k}\Im R({\bf k}\omega)]^2},
\end{equation}
where $f_{\bf k}=(i\dot{s}^z_{\bf k},-i\dot{s}^z_{\bf -k})^{-1}$,
$(A,B)=i\int_0^\infty dt\langle[A(t),B]\rangle$, the angular brackets
denote the statistical averaging, $A(t)=e^{iHt}Ae^{-iHt}$ with the
Hamiltonian $H$, $\mu_B$ is the Bohr magneton, $s^z_{\bf
k}=N^{-1/2}\sum_{\bf n}e^{i{\bf kn}}s^z_{\bf n}$ with the number of
sites $N$ and the $z$ component of the spin $s^z_{\bf n}$ on the
lattice site {\bf n}, $i\dot{s}^z_{\bf k}=[s^z_{\bf k},H]$,
$\omega^2_{\bf k}=(i\dot{s}^z_{\bf k},-i\dot{s}^z_{\bf -k})(s^z_{\bf
k},s^z_{\bf -k})^{-1}$, $R({\bf k}\omega)=-i\int_0^\infty
dt\,e^{i\omega t}(A_{2t},A_2^\dagger)$ is the memory function with
$A_2=i^2\ddot{s}^z_{\bf k}-\omega^2_{\bf k}s^z_{\bf k}$,
$$i\frac{d}{dt}A_{2t}=(1-P_0)(1-P_1)[A_{2t},H],\; A_{2,t=0}=A_2,$$
and the projection operators $P_k$, $k=0,1$, are defined as
$$P_kQ=(Q,A_k^\dagger) (A_k,A_k^\dagger)^{-1}A_k$$ with the operators
$A_0=s^z_{\bf k}$ and $A_1=i\dot{s}^z_{\bf k}$.

To describe the spin excitations of Cu-O planes which determine the
magnetic properties of cuprates the $t$-$J$ model \cite{Izyumov} is
used. Its Hamiltonian reads
\begin{equation}\label{hamiltonian}
H=\sum_{\bf nm\sigma}t_{\bf nm}a^\dagger_{\bf n\sigma}a_{\bf
m\sigma}+\frac{1}{2}\sum_{\bf nm}J_{\bf nm}{\bf s_n s_m},
\end{equation}
where $a_{\bf n\sigma}=|{\bf n}\sigma\rangle\langle{\bf n}0|$ is the
hole annihilation operator, {\bf n} and {\bf m} label sites of the
square lattice, $\sigma=\pm 1$ is the spin projection, $|{\bf
n}\sigma\rangle$ and $|{\bf n}0\rangle$ are site states corresponding
to the absence and presence of a hole on the site, $s^z_{\bf
n}=\frac{1}{2}\sum_\sigma\sigma|{\bf n}\sigma\rangle\langle{\bf
n}\sigma|$ and $s^\sigma_{\bf n}=|{\bf n}\sigma\rangle\langle{\bf
n},-\sigma|$ are the spin-$\frac{1}{2}$ operators. Considering the
interaction between nearest neighbor spins, described by the parameter
$J$, and the hole hopping to nearest and next nearest neighbor sites,
described by the parameters $t$ and $t'$, respectively, the terms in
Eq.~(\ref{chi}) are brought to the form \cite{Sherman02}
\begin{eqnarray}
&&(i\dot{s}^z_{\bf k},-i\dot{s}^z_{\bf -k})=4(1-\gamma_{\bf k})
 (J|C_1|+tF_1), \nonumber\\
&&\omega_{\bf k}^2=16J^2\alpha|C_1|(1-\gamma_{\bf k})(\Delta+1+
 \gamma_{\bf k}), \nonumber\\[-0.5em]
&&\label{chitJ}\\[-0.5em]
&&\Im R({\bf k}\omega)=\frac{8\pi t\omega^2_{\bf k}}{N}
 \sum_{\bf k'}g_{\bf kk'}^2\int_{-\infty}^\infty d\omega' A({\bf
 k'}\omega') \nonumber\\
&&\hspace{4em}\times A({\bf k+k'},\omega+\omega')
 \frac{n_F(\omega+\omega')-n_F(\omega')}{\omega},\nonumber
\end{eqnarray}
where $\gamma_{\bf k}=\frac{1}{2}[\cos(k_x)+\cos(k_y)]$, $C_1=\langle
s^{+1}_{\bf n}s^{-1}_{\bf n+a}\rangle$ and $F_1=\langle a^\dagger_{\bf
n}a_{\bf n+a}\rangle$ are the spin and hole correlations on neighbor
sites, {\bf a} is a vector connecting these sites, $\alpha\sim 1$ is
the vertex correction parameter, $g_{\bf kk'}=\gamma_{\bf
k'}-\gamma_{\bf k+k'}+\frac{t'}{t}(\gamma'_{\bf k'}-\gamma'_{\bf
k+k'})$, $\gamma'_{\bf k}=\cos(k_x)\cos(k_y)$,
$n_F(\omega)=[\exp(\omega/T)+1]^{-1}$ with the temperature $T$, $A({\bf
k}\omega)$ is the hole spectral function which is taken in the form
\begin{equation}\label{hsf}
A({\bf k}\omega)=\frac{\eta/\pi}{(\omega-\varepsilon_{\bf
k}+\mu)^2+\eta^2}
\end{equation}
in this work. Here $\mu$ is the chemical potential, $\eta$ is the
artificial broadening, and $\varepsilon_{\bf k}$ is the hole
dispersion.

In Eq.~(\ref{chitJ}), $\Delta\propto\xi^{-2}$ where $\xi$ is the
correlation length of the short-range antiferromagnetic order
\cite{Sherman03}. Thus, in the short-range order the frequency of spin
excitations at {\bf Q} is nonzero, in contrast to the classical
antiferromagnetic magnons. As follows from Eq.~(\ref{chitJ}), the
dispersion of spin excitations has a local minimum at {\bf Q} and can
be approximated as $\omega_{\bf k}=[\omega^2_{\bf Q}+c^2({\bf
k-Q})^2]^{1/2}$ near this momentum \cite{Sherman02}.

To simplify calculations let us set $T=0$. Notice that $\Im R({\bf
k}\omega)$ is an even function of the frequency and consider the region
$\omega\geq 0$. The integral in Eq.~(\ref{chitJ}) reduces to
$\int_{-\omega}^0d\omega'\,A({\bf k'}\omega') A({\bf
k+k'},\omega+\omega')$ and for the spectral function (\ref{hsf}) is
easily integrated. For $\eta\ll\omega$ the states with energies
$-\omega<\varepsilon_{\bf k'}-\mu<0$ and $0<\varepsilon_{\bf
k+k'}-\mu<\omega$ make the main contribution to this integral.

In the following, we use the values of $C_1$, $F_1$, $\Delta$ and
$\alpha$ calculated self-con\-sis\-ten\-t\-ly in the $t$-$J$ model on a
20$\times$20 lattice for the range of hole concentrations $0\leq
x\lesssim 0.16$ \cite{Sherman03}. The calculations were carried out for
the parameters $t=0.5$~eV and $J=0.1$~eV corresponding to hole-doped
cuprates \cite{McMahan}. For $\varepsilon_{\bf k}$ we apply the hole
dispersion $\varepsilon_{\bf k}=-0.0879+0.5547\gamma_{\bf
k}-0.1327\gamma'_{\bf k}-0.0132\gamma_{2\bf k}
+0.0925[\cos(2k_x)\cos(k_y)+ \cos(k_x)\cos(2k_y)]-0.0265\gamma'_{2\bf
k}$ proposed from the analysis of photoemission data in
Bi$_2$Sr$_2$CaCu$_2$O$_8$ \cite{Norman}. Here the coefficients are in
electronvolts. Analogous results can be obtained also with other model
dispersions \cite{Liu,Norman}.

\begin{figure}[t]
\centerline{\includegraphics[width=6.5cm]{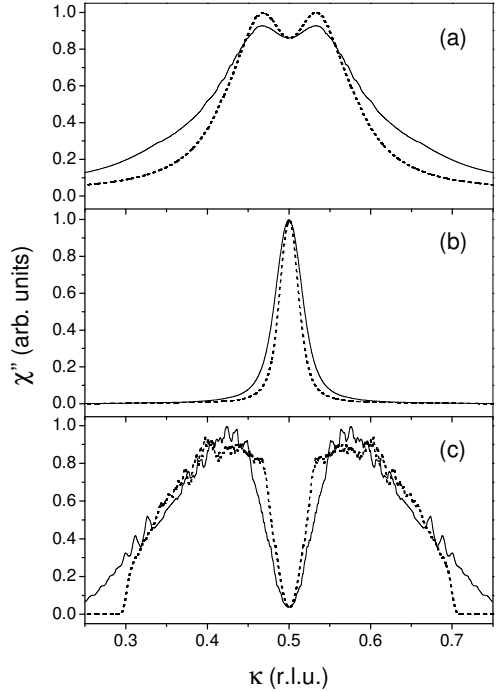}}
\caption{Wave-vector scan of $\chi''$ for $T=0$, $x \approx 0.12$,
$\mu=-40$~meV, and $\omega=70$~meV, $\eta=3.5$~meV (a),
$\omega=35$~meV, $\eta=3.5$~meV (b), $\omega=2$~meV, $\eta=1.5$~meV
(c). Calculations were carried out in a 1200$\times$1200 lattice. The
solid lines correspond to scans along the edge of the Brillouin zone,
${\bf k}=(\kappa,\frac{1}{2})$; the dashed lines are for the zone
diagonal, ${\bf k}=(\kappa,\kappa)$.} \label{Fig_i}
\end{figure}

\begin{figure}[t]
\centerline{\includegraphics[width=7cm]{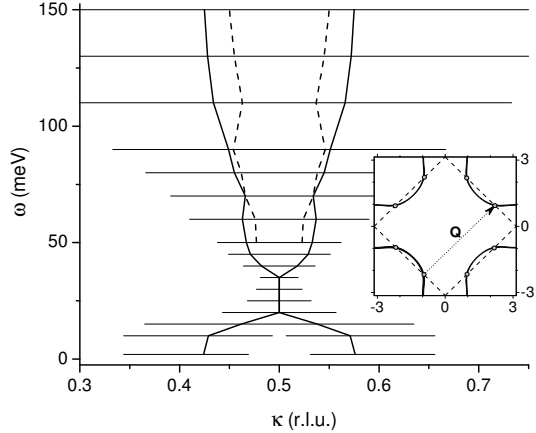}} \caption{The
dispersion of maxima in $\chi''({\bf k}\omega)$ for scans along the
edge [${\bf k}=(\kappa,\frac{1}{2})$, solid lines] and the diagonal
[${\bf k}=(\kappa,\kappa)$, dashed lines] of the Brillouin zone. The
latter dispersion are drawn only in the frequency range in which these
maxima are more intensive than those along the edge. Parameters are the
same as in Fig.~\protect\ref{Fig_i}. Horizontal bars are FWHM for
maxima along the edge. Inset: The Fermi surface for $\mu=-40$~meV (the
solid line). Dashed lines show the boundary of the magnetic Brillouin
zone, circles are the hot spots.} \label{Fig_ii}
\end{figure}

The momentum dependence of $\chi''({\bf k}\omega)$ calculated with the
above equations is shown in Fig.~\ref{Fig_i}.
As seen from this figure, there are three frequency regions with
different shapes of the momentum dependence of $\chi''({\bf k}\omega)$.
The first region is the vicinity of the frequency $\omega_{\bf Q}$ of
the gap in the dispersion of spin excitations at {\bf Q}. For the
parameters of Fig.~\ref{Fig_i} $\omega_{\bf Q}\approx 37$~meV. In this
region $\chi''$ is peaked at {\bf Q}. For smaller and larger
frequencies the magnetic response is incommensurate. The dispersion of
maxima in $\chi''({\bf k}\omega)$ for scans along the edge and the
diagonal of the Brillouin zone and their full widths at half maximum
(FWHM) are shown in Fig.~\ref{Fig_ii}.

To understand the above results notice that $\chi''$ in Eq.~(\ref{chi})
has the resonance denominator which will dominate in the momentum
dependence for $\omega \geq\omega_{\bf Q}$ if the spin excitations are
not overdamped. Parameters of Fig.~\ref{Fig_i} correspond to this case.
For $\omega \geq\omega_{\bf Q}$ the equation $\omega=\omega_{\bf k}$
determines the positions of the maxima in $\chi''({\bf k}\omega)$ which
are somewhat shifted by the momentum dependence of $\Im R$ (for
simplicity the term with $\Re R$ is included in $\omega_{\bf k}$).
Using the above approximation for $\omega_{\bf k}$ we find that the
maxima in $\chi''$ are located near a circle centered at {\bf Q} with
the radius $c^{-1}(\omega^2-\omega^2_{\bf Q})^{1/2}$
\cite{Sherman02,Barzykin}.

For $\omega<\omega_{\bf Q}$ the nature of incommensurability is
completely different. It is most easily seen in the limit of small
$\omega$ when Eq.~(\ref{chi}) reduces to $\chi''({\bf
k}\omega)\propto\omega_{\bf k}^{-4}\Im R({\bf k}\omega)$.
$\omega^{-4}_{\bf k}$ is a decreasing function of the difference ${\bf
k-Q}$ which acts in favor of a commensurate peak. However, if $\Im
R({\bf k}\omega)$ has a pronounced dip at {\bf Q} incommensurate peaks
arise. For small $x$ the dip appears due to the nesting of the ellipses
forming the Fermi surface with the wave vector {\bf Q}
\cite{Sherman04}. For larger $x$ the mechanism of the dip formation is
the following. For ${\bf k=Q}$ and small $\omega$ hole states which
make the main contribution to $\Im R$ are located near the hot spots
(see the inset in Fig.~\ref{Fig_ii}). For these states the interaction
constant $g_{\bf Qk'}$ in Eq.~(\ref{chitJ}) is small which leads to the
smallness of $\Im R$. With moving away from {\bf Q} momenta of states
contributing to the damping recede from the hot spots and $g_{\bf kk'}$
and $\Im R$ grow. Thus, the damping has a dip at {\bf Q} which leads to
the low-frequency incommensurability shown in Fig.~\ref{Fig_i}c. The
smallness of $g_{\bf Qk'}$ near the hot spots is related to the fact
that the decaying site spin excitation creates the fermion pair on the
same and neighbor sites. As a result $g_{\bf kk'}$ contains the
functions $\gamma_{\bf k'}$, $\gamma_{\bf k+k'}$ and $\gamma'_{\bf
k'}-\gamma'_{\bf k'+k}$ vanishing on the boundary of the magnetic
Brillouin zone and at ${\bf k=Q}$.

\begin{figure}[t]
\centerline{\includegraphics[width=6.5cm]{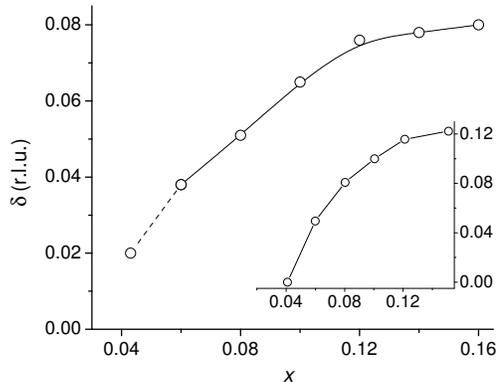}} \caption{The
incommensurability parameter $\delta$ vs.\ $x$ for $\omega=2$~meV. The
value of $\delta$ for $x=0.043$ was taken from
Ref.~\protect\cite{Sherman04}. Inset: experimental data
\protect\cite{Yamada} for La$_{2-x}$Sr$_x$CuO$_4$. Connecting lines are
a guide to the eye.} \label{Fig_iii}
\end{figure}

The dispersion of peaks in $\chi''$ which is similar to that shown in
Fig.~\ref{Fig_ii} was observed in YBa$_2$Cu$_3$O$_{7-y}$ and
La$_{2-x}$Ba$_x$CuO$_4$ \cite{Dai,Tranquada}. As seen from
Fig.~\ref{Fig_i}, for frequencies $\omega<\omega_{\bf Q}$ the
susceptibility is peaked at ${\bf
k}=(\frac{1}{2},\frac{1}{2}\pm\delta)$ and
$(\frac{1}{2}\pm\delta,\frac{1}{2})$, while for $\omega>\omega_{\bf Q}$
the maxima are located at
$(\frac{1}{2}\pm\delta,\frac{1}{2}\pm\delta)$,
$(\frac{1}{2}\pm\delta,\frac{1}{2}\mp\delta)$. This result is also in
agreement with experimental observations \cite{Dai,Hayden04,Tranquada}.
Notice that in these crystals Fermi surfaces are similar to that shown
in Fig.~\ref{Fig_ii} \cite{Damascelli}. The dependence of the
incommensurability parameter $\delta$ on $x$ for small $\omega$ is
shown in Fig.~\ref{Fig_iii}.
In agreement with experiment (see the inset in Fig.~\ref{Fig_iii})
$\delta$ grows nearly linearly with $x$ up to $x\lesssim 0.12$ and then
saturates. In these calculations the change of the hole dispersion with
doping was not considered and the dependence $\delta(x)$ is connected
solely with the variation of $\omega_{\bf Q}\propto\xi^{-1}\propto
x^{1/2}$ \cite{Sherman03}. We found that the low-frequency
incommensurability disappears when $\eta>\omega$. Besides, it
disappears also if $\mu$ approaches the extended van Hove singularities
at $(0,\pi)$, $(\pi,0)$. In this case for ${\bf k=Q}$ the entire region
of these singularities in which $g_{\bf kk'}$ is nonzero contributes to
$\Im R$. This may be the reason of the disappearance of the
incommensurability in overdoped cuprates \cite{Yamada}. In the
calculations the incommensurability disappears also in small lattices.

\begin{figure}[t]
\centerline{\includegraphics[width=6.5cm]{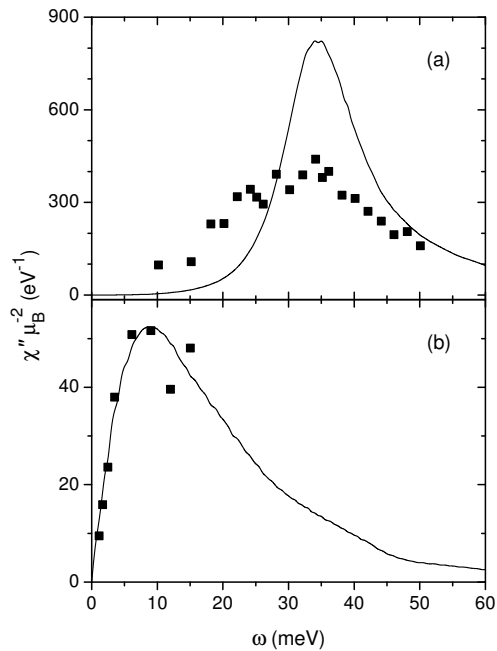}} \caption{The
frequency dependence of $\chi''$. The solid lines are our results for
$T=0$, $x \approx 0.12$, $\mu=-40$~meV, $\eta=3.5$~meV, ${\bf k=Q}$ (a)
and for ${\bf k}=(0.42,0.5)$ and the scaled hole dispersion (see text)
(b). Squares are the odd susceptibility in the normal-state
YBa$_2$Cu$_3$O$_{6.83}$ ($x\approx 0.14$ \protect\cite{Tallon}) at
$T=100$~K and ${\bf k=Q}$ \protect\cite{Bourges} (a) and the
susceptibility in La$_{1.86}$Sr$_{0.14}$CuO$_4$ for $T=35$~K at the
incommensurate peak \protect\cite{Aeppli} (b).} \label{Fig_iv}
\end{figure}
As mentioned above, for the parameters chosen spin excitations are not
overdamped near {\bf Q}. As a consequence the frequency dependence of
$\chi''({\bf Q}\omega)$ has a pronounced maximum at
$\omega\approx\omega_{\bf Q}$ which resembles the susceptibility
observed in the superconducting and normal states of underdoped
YBa$_2$Cu$_3$O$_{7-y}$. As seen in Fig.~\ref{Fig_iv}a, the experimental
width of the maximum is approximately twice as large as the calculated
one. Partly it is connected with the difference in temperatures for the
two sets of the data. Besides, the decrease of the hole bandwidth and
the increase of the hole damping lead to a substantial growth of $\Im
R$ and to the overdamping of spin excitations \cite{Sherman02}. An
example of such changes is shown in Fig.~\ref{Fig_iv}b where the
results were obtained with the hole dispersion scaled by the factor
0.4. The calculations were carried out for ${\bf k}=(0.42,0.5)$ which
corresponds to the low-frequency incommensurate peak in
Fig.~\ref{Fig_i}c. The growth of $\Im R$ leads to the red shift of the
maximum in $\chi''(\omega)$. Its position is no longer connected with
the frequency of spin excitations. The similar frequency dependence of
$\chi''$ is observed in La$_{2-x}$Sr$_x$CuO$_4$ \cite{Aeppli}. The
increased $\Im R$ has no marked effect on the low-frequency
incommensurability, however for the frequencies $\omega_{\bf
Q}\geq\omega\geq 150$~meV we found only broad commensurate peaks for
the parameters of Fig.~\ref{Fig_iv}b.

Our consideration was restricted to the normal state. Clearly the same
mechanisms lead to the incommensurate magnetic response also in the
superconducting state.

In summary, it was shown that the calculated momentum and frequency
dependencies of the imaginary part of the susceptibility $\chi''$, the
dispersion and location of maxima in it and the concentration
dependence of the incommensurability parameter are similar to those
observed in lanthanum and yttrium cuprates. The dispersion of the
maxima in $\chi''$ resembles two parabolas with upward and downward
branches which converge at the antiferromagnetic wave vector {\bf Q}
and at the respective frequency of spin excitations. We relate the
upper parabola to the spin-excitation dispersion. The
incommensurability connected with the lower parabola is related to the
dip in the spin-excitation damping at {\bf Q}. For moderate doping the
dip arises due to the smallness of the interaction between spin
excitations and holes near the hot spots. In agreement with experiment
the incommensurate peaks which form the lower parabola are located at
momenta $(\frac{1}{2},\frac{1}{2}\pm\delta)$,
$(\frac{1}{2}\pm\delta,\frac{1}{2})$, while peaks in the upper parabola
are at $(\frac{1}{2}\pm\delta,\frac{1}{2}\pm\delta)$,
$(\frac{1}{2}\pm\delta,\frac{1}{2}\mp\delta)$. Also in agreement with
experiment the low-frequency incommensurability parameter $\delta$
grows linearly with the hole concentration $x$ for $x\lesssim 0.12$ and
then saturates. By the variation of the hole bandwidth and damping the
frequency dependencies of the susceptibility which resemble those
observed in YBa$_2$Cu$_3$O$_{7-y}$ and La$_{2-x}$Sr$_x$CuO$_4$ were
obtained.

This work was supported by the ESF grant No.~5548.

\end{document}